\documentclass[a4paper]{spie}  

\usepackage{amsmath,amsfonts,amssymb}
\usepackage{graphicx}
\usepackage[colorlinks=true, allcolors=blue]{hyperref}

\def\apj{ApJ}                 
\def\aap{A\&A}                
\def\mnras{MNRAS}             

\title{Scientific simulations and optimization of the XGIS instrument on board THESEUS}

\author[a]{S.~Mereghetti}
\author[c]{G.~Ghirlanda}
\author[a]{R.~Salvaterra}
\author[b]{R.~Campana}
\author[b]{C.~Labanti}
\author[e]{P.H.~Connell}
\author[b]{R.~Farinelli}
\author[d]{F.~Frontera}
\author[b]{F.~Fuschino}
\author[e]{J.L.~Gasent-Blesa}
\author[d]{C.~Guidorzi}
\author[a]{M.~Lissoni}
\author[a]{M.~Rigoselli}
\author[b]{J.B.~Stephen}
\author[b]{L.~Amati}

\affil[a]{INAF / IASF-Milano, via A. Corti 12, I-20133 Milano, Italy}
\affil[b]{INAF/OAS-Bologna, via P. Gobetti 101, I-40129 Bologna, Italy}
\affil[c]{INAF/Osservatorio Astronomico di Brera, via E.  Bianchi 46, I-23807 Merate, Italy}
\affil[d]{Università di Ferrara and INFN, via Saragat 1, I-44122 Ferrara, Italy}
\affil[e]{Image Processing Laboratory, University of Valencia, c/ Catedrático José Beltrán, 2, E46980, Paterna (Valencia), Spain}
 
\authorinfo{
Send correspondence to Sandro Mereghetti (sandro.mereghetti@inaf.it) 
}

\begin{document} 
\maketitle

\begin{abstract}
The XGIS (X and Gamma Imaging Spectrometer) is one of the three instruments onboard the THESEUS mission (ESA M5, currently in Phase-A). Thanks to its wide field of view and good imaging capabilities, it will efficiently detect and localize gamma-ray bursts and other transients in the 2-150 keV sky, and also provide spectroscopy up to 10 MeV. Its current design has been optimized by means of scientific simulations based on a Monte Carlo model of the instrument coupled to a state-of-the-art description of the populations of long and short GRBs extending to high redshifts. We describe the optimization process that led to the current design of the XGIS, based on two identical units with partially overlapping fields of view, and discuss the expected performance of the instrument.
\end{abstract}


\section{INTRODUCTION}
\label{sec:intro}  

THESEUS (Transient High Energy Sky and Early Universe Surveyor)  is a  proposed space mission with the two main scientific objectives of  {\it a)} exploring the early Universe through the detection and characterization of  gamma-ray bursts (GRBs) at high redshift and {\it b)} identifying  the electromagnetic counterparts of gravitational waves sources \cite{2018AdSpR..62..191A}. THESEUS has been selected by the European Space Agency in 2018 for an assessment study in response  to the call for the 5$^{th}$ Medium Size mission of the Cosmic Vision program, with  a launch foreseen in 2032. 

Its payload comprises wide field X-ray and $\gamma$-ray monitors, capable of discovering and accurately localizing GRBs and other transient high-energy sources, and an infrared telescope for their  characterization with rapid follow-up observations after  autonomous slews of the satellite.  
These  instruments   also allow THESEUS to obtain extremely valuable X-ray and  $\gamma$-ray data for time-domain studies of a large variety of high-energy sources, as well as to exploit the infrared telescope for  Guest Observer observations that can be carried out in parallel with the main THESEUS science program.  

More specifically, the THESEUS payload will comprise: 

\begin{itemize}

\item  a Soft X-ray Imager (SXI\cite{obrien20}) operating  in the 0.3-5 keV energy range and with  a field of view of $\sim$30$\times$60 deg$^2$, thanks to  CMOS focal plane detectors and micropore optics in lobster-eye configuration;

\item a X-Gamma-rays Imaging Spectrometer (XGIS\cite{labanti20}), consisting of two identical coded mask telescopes,  which provides a wide and deep sky monitoring in the broad energy band from  2 keV to 10 MeV;

\item an Infrared Telescope (IRT\cite{gotz20}) with 70 cm aperture, providing both imaging and spectroscopy (R$\sim$400) capabilities in the 0.7-1.8 $\mu$m range;

\end{itemize}

THESEUS will be placed in a low equatorial orbit (height $\sim$600 km, inclination $<6^{\circ}$), ensuring a low and stable instrumental background level, and will operate with a pointing strategy optimized to maximise the number of detected GRBs and, at the same time,  facilitate  their follow-up  with large ground based telescopes.  The sky coordinates and other information on the detected GBRs will be transmitted to ground in  real time, through a dedicated network of VHF ground stations.  If selected for the subsequent phases, THESEUS will be launched in 2032 and have a nominal lifetime of four years.

In this paper we  concentrate on  the XGIS  instrument, and in particular on its high level scientific requirements, on the optimization process that led to the currently proposed configuration, and on its predicted performances.

   \begin{figure} [ht]
   \begin{center}
   \begin{tabular}{cc} 
   \includegraphics[height=7cm]{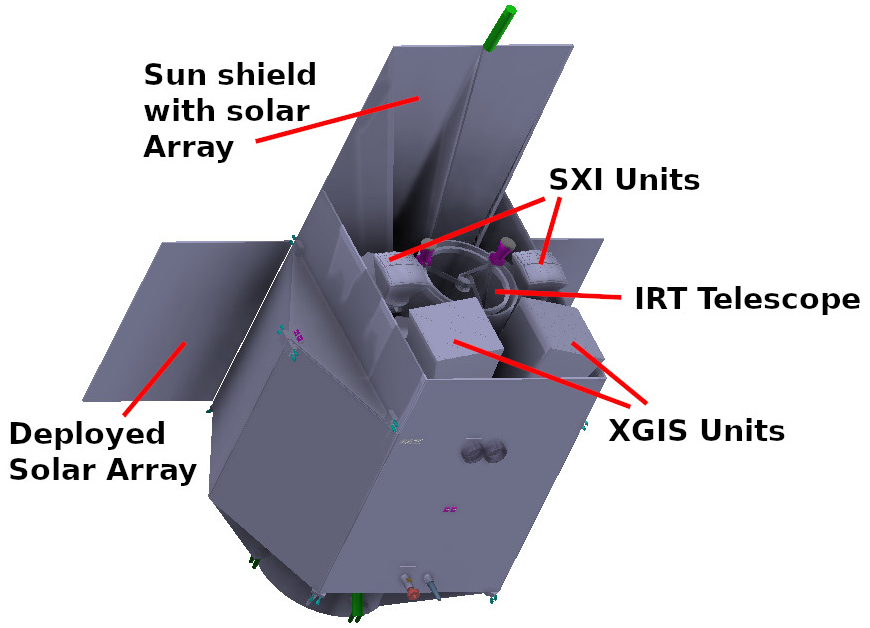}
      \includegraphics[height=7cm]{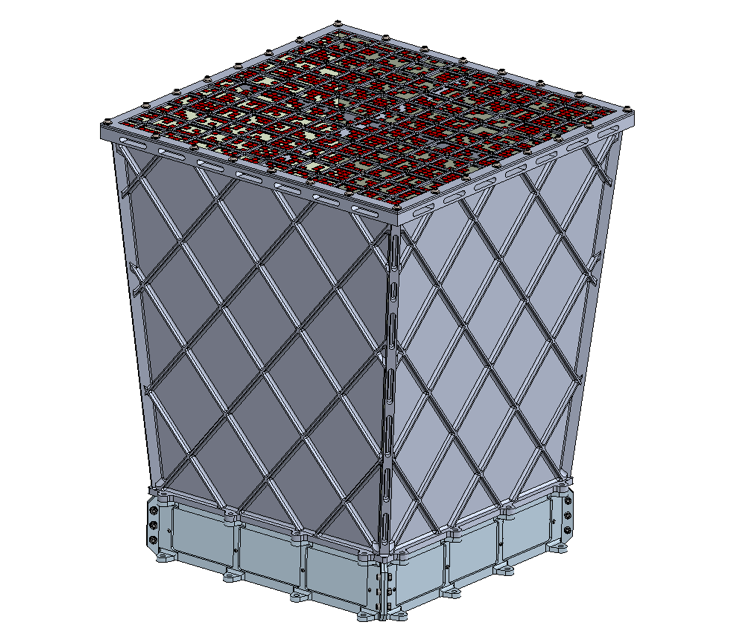}
   \end{tabular}
   \end{center}
   \caption[THESEUS satellite concept] 
   { \label{fig:theseus} 
Left:  concept of the THESEUS satellite. Note the different pointing directions of the two XGIS cameras.   Right: one of the two identical XGIS cameras. Its dimensions are $\sim$60$\times$60$\times$91 cm$^3$ and its weight is about 80 kg.   }
   \end{figure}

\section{SCIENCE REQUIREMENTS AND DESIGN TRADE-OFFS}
\label{sec:req}  

GRBs arrive at unpredictable times from random directions in the sky. To a large extent, this is also true for most impulsive sources of gravitational waves, such as those produced in the coalescence of  neutron stars and/or black holes in compact binary systems.  Furthermore, GRBs are  bright only for short time intervals and then they fade rapidly.  Therefore, the capability  to monitor a large fraction of the sky is  a fundamental requirement for any satellite aiming to  detect and study these phenomena.  A good localization accuracy is also necessary, in order to perform multiwavelength follow-ups, and this must be obtained as soon as possible.   A broad energy range is needed to characterize the GRBs spectral shape, in particular to properly measure the energy at which the emission peaks  (a fundamental parameter for cosmological applications of GRBs). An operating range extending to  low energies favours the detection of GRBs at high redshift \cite{ghi15}, that is one of the main objectives of THESEUS.  Good time and energy resolutions are also important aspects to consider in the choice of the detectors. 

The  imaging-related aspects of these requirements are best achieved in the hard X-ray range thanks to instruments based on the  coded mask technique, as it has been successfully demonstrated by many past and current missions \cite{}.  Coded mask imaging necessitates of  a position-sensitive detector able to record the shadows of the aperture pattern cast by the sources in the field of view. The recorded ``shadowgrams'' can then be processed to reconstruct images of the sky. A detector made of individual pixels arranged in a bidimensional array provides this capability.
In the XGIS the individual detector elements are small scintillator bars made of CsI(Tl) coupled with Silicon Drift Detectors (SDD).  The SDDs (two for each bar) are used to collect the scintillation light produced in the CsI by interacting photons with energy above $\sim$30 keV, and also for the direct detections of lower energy photons, down to a threshold of 2 keV.  Using these basic elements as detector pixels, it is possible to build, in a modular way, a detector with 3-D location capability\footnote{Localization along the CsI bar is obtained by comparing the signals from the two SDD.},  large effective area, and wide energy bandwidth from 2 keV to 10 MeV. 

The XGIS instrument conceptual design has been driven by the need to achieve the THESEUS science requirements taking into account the  payload resources for a Medium Size satellite mission (e.g., mass, power, dimensions, telemetry), the inevitable trade-offs between contrasting design aspects (e.g., field of view, sensitivity and angular resolution), as well as the need of a high level of technological readiness in order  to ensure a reliable and timely development phase.

Taking advantage of the modular design of the detection elements, we explored different  configurations based either on a single  telescope with a rectangular detection plane or on a few smaller square telescopes, arranged in such a way as to provide a similar sky coverage.  The comparison between the different configurations was done by estimating the  expected number and redshift distribution  of the detectable GRBs, taking into account the energy and angular dependence of the instrumental sensitivity.  
The expected number of detectable GRBs were derived with  simulations based on a  model of the instrument and  of  the   populations of GRBs of the long and short classes as described in Section \ref{sec:pop}. The instrumental background was derived through detailed Monte Carlo simulations \cite{campana20}. It must be noted that the dominant source of background  in the lower energy range of the XGIS is given by the diffuse X-ray emission of cosmic origin, which increases with the dimensions of the field of view.  Based on the estimated number of GRBs (and on their redshift distribution) and considering payload accommodation constraints, a design consisting of two identical units (``cameras'') with partially overlapping fields of view was adopted.

The geometry of the cameras was optimized starting from the minimum size of the pixels that can be reliably achieved with the proposed detector technology, and exploring different  trade-offs between field of view and sensitivity. The coded mask properties and those of the collimating system were designed to allow imaging up to energies of $\sim$150 keV, in order to increase the sensitivity for short GRBs, which have a hard spectrum and are particularly interesting in the context of counterparts of gravitational waves sources.  Further design optimizations will be carried out during Phase-B, if the mission is selected.

\section{INSTRUMENT DESCRIPTION}
\label{sec:instr}  

The XGIS is composed of two identical cameras, that, in the baseline THESEUS payload configuration studied in Phase-A, will be pointed at directions offset by $\pm20^{\circ}$ from the satellite boresight direction (that coincides with the pointing direction of the IRT, see left panel of Fig.~\ref{fig:theseus}).  In this section we briefly describe the design of an XGIS camera (Fig.~\ref{fig:theseus}, right panel).

The detection plane is an array of  80$\times$80 pixels, grouped in 100 modules of 64 pixels. Each pixel consists of a CsI(Tl) scintillator bar, with height of 30 mm and square cross section of 4.5$\times$4.5 mm$^2$, sensitive in the nominal 30 keV-10 MeV  energy range. The SDD on the top of each bar, besides being  used as a readout for the scintillator light, acts as a detector for soft X-ray photons. Its thickness of 450 $\mu$m provides a high efficiency  in the nominal 2-30 keV  energy range. The modules composing the detection plane have a pitch, defined as the distance between the centers of adjacent pixels, of 5 mm.  The space between modules, required for mechanical and signal read out reasons, has a width equal to the pitch. Thus, from the point of view of the coded mask imaging, the detector array can be considered as a matrix of 89$\times$89 elements (including nine ``dead'' rows and nine ``dead'' columns between the modules) with overall dimensions of 44.5$\times$44.5 cm$^2$.

The coded mask is an array of square elements, made of tungsten with thickness of 1 mm, placed at a distance of 63 cm from the top surface of the detector\cite{gasent20}. The  dimensions of  the mask pattern are $\sim$57$\times$57 cm$^2$.  The mask is supported by a mechanical structure that connects it to the detector and also acts as a passive shield to delimit the field of view at low energies.  The mask pattern design and the dimensions of the mask elements will be optimized during the next phases of the project. 

The XGIS instrument also comprises two Power Supply Units (one for each camera) and a Data Handling Unit. The latter, in addition to all the standard functions related to instrument control and data management, has the fundamental task of running the on-board software for the GRB trigger and localization using the data of the two XGIS cameras (as well as complementary information from the SXI instrument and from the spacecraft).

   \begin{table}[ht]
\caption{XGIS Scientific Performances  } 
\label{tab:Multimedia-Specifications}
\begin{center}       
\begin{tabular}{ | p{4cm} |p{8cm} |p{4cm} |}
\hline
\rule[-1ex]{0pt}{3.5ex} 
   & Coded mask imaging performances & Wide FoV non-imaging performances  \\
\hline\rule[-1ex]{0pt}{3.5ex} 
 Energy range & 2-150 keV & 150 keV - 10 MeV  \\
 
\hline\rule[-1ex]{0pt}{3.5ex} 
 Field of view & 10$\times$10 deg$^2$  fully coded, one camera &  4 sr ($>$20\% efficiency)\\
                     & 77$\times$77 deg$^2$  zero sensitivity, one camera &    \\
                     & 80$\times$45 deg$^2$  half  effective area, combined cameras &    \\
                    & 117$\times$77 deg$^2$  zero  sensitivity, combined cameras &    \\
                    
\hline\rule[-1ex]{0pt}{3.5ex} 
 Sensitivity  (3$\sigma$ in 1 s) & $10^{-8}$ erg cm$^{-2}$ s$^{-1}$ (2-30 keV)  & $2.7\times10^{-7}$ erg cm$^{-2}$ s$^{-1}$  (0.15-1 MeV)     \\
                                               & $3\times10^{-8}$ erg cm$^{-2}$ s$^{-1}$  (30-150 keV) &    \\

\hline\rule[-1ex]{0pt}{3.5ex} 
 Point spread function & $\sim$1 deg (FWHM)  &   -  \\
 
\hline\rule[-1ex]{0pt}{3.5ex} 
 Source location  accuracy (90\% c.l. error radius) &   $<$15 arcmin (for a source with $SNR>$9)  &   - \\

\hline\rule[-1ex]{0pt}{3.5ex} 
Energy resolution &   $<$ 1200 eV  (FWHM at 6 keV)  &    6\% (FWHM at 500 keV) \\

\hline\rule[-1ex]{0pt}{3.5ex} 
Relative timing accuracy  & \multicolumn{2}{c|}{7$\mu$s}    \\
\hline

\end{tabular}
\end{center}
\end{table}

\section{SCIENTIFIC PERFORMANCES}
\label{sec:perf}  

The XGIS has been conceived for two complementary aspects: {\it i)} imaging, for source identification and accurate localization and  {\it ii)} spectroscopy, for source characterization over a broad energy range. The first aspect is best achieved in the lower energy range, exploiting the coded mask   in an imaging field of view  defined by the mask and detector geometry, 
as described below. On the other hand, in the higher part of the energy range, the absorbing power of the closed  mask elements is too small to significantly modulate the incoming radiation and  the passive shielding, that limits the imaging field of view, becomes increasingly transparent with energy. As a result, the XGIS can detect high-energy radiation also from sources outside the imaging  field of view. For such sources,  only some limited positional information can be derived (e.g., by comparing the count rates in the two cameras), but full timing and spectral information will be obtained. For simplicity, the values  listed in Table 1   refer to the imaging and non-imaging performances  with an energy division at 150 keV,  but it should be kept in mind that this is just a conventional energy threshold and, in reality, there is a gradual transition between the two regimes.

\subsection{Imaging field of view}
\label{sec:fov}  

\begin{figure} [ht]
   \begin{center}
   \begin{tabular}{cc} 
    \includegraphics[height=6cm]{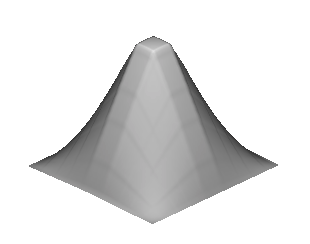}
         \includegraphics[angle=90,height=6cm]{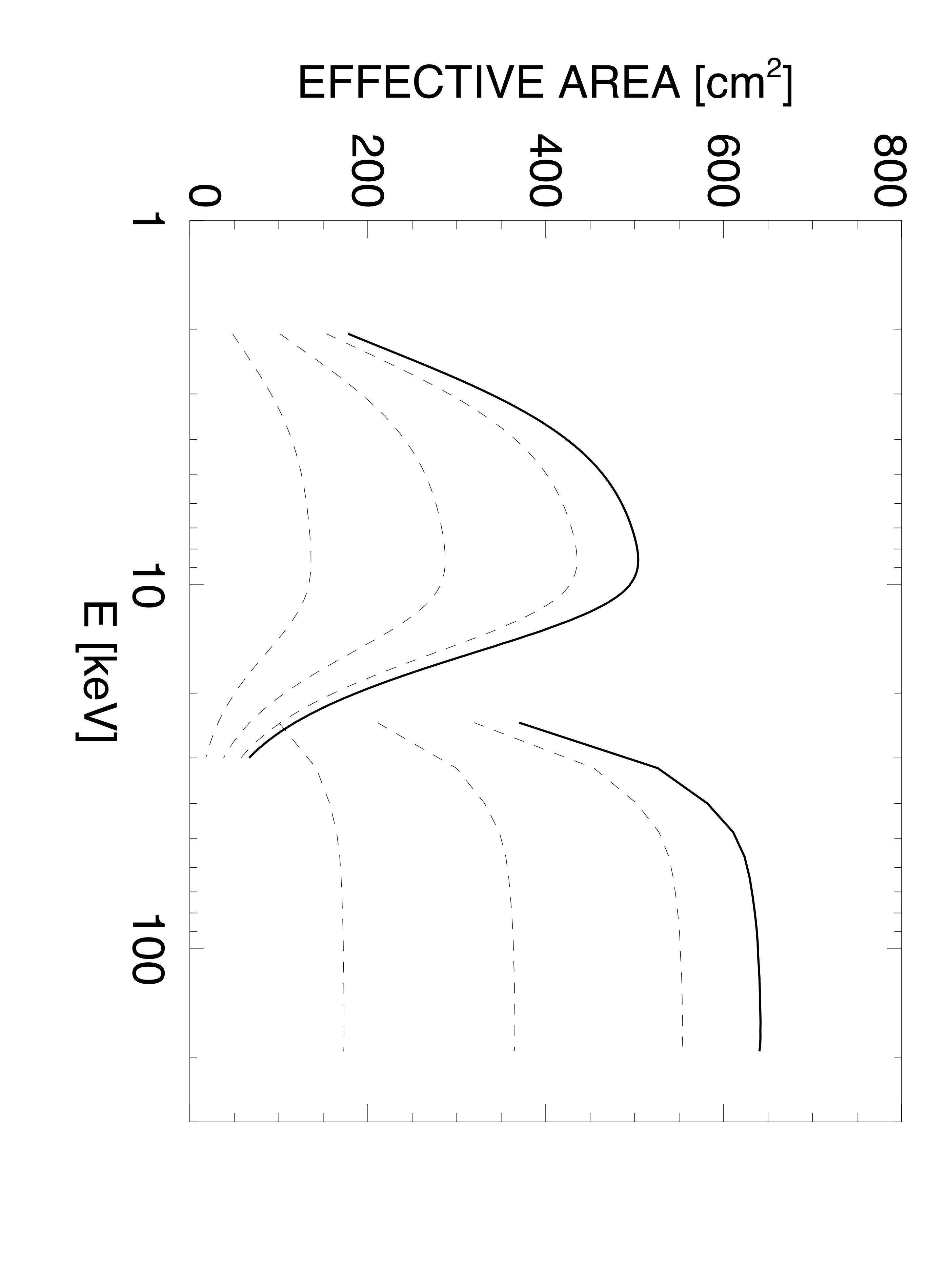}    
   \end{tabular}
   \end{center}
   \caption[THESEUS satellite concept] 
   { \label{fig:fov} 
Left:  Angular dependence of the detector working area (i.e. the detector geometric surface that records a source flux modulated by the mask pattern) for one XGIS camera. The top value is achieved in the Fully Coded FoV of $\sim10^{\circ}\times10^{\circ}$.  The zero response has a size of $77^{\circ}\times77^{\circ}$ deg$^2$. 
Right: Effective area (one unit) as a function of energy. The top curves are for the FCFoV (off-axis angles $<$5$^{\circ}$). The lower curves are for off-axis angles of 10$^{\circ}$, 20$^{\circ}$,  and 30$^{\circ}$.  }
   \end{figure}

In coded mask imaging systems, the field of view (FoV) is defined by the   dimensions of     the mask ($D_M$), of the   detector ($D_D$), and their distance ($H$).  The central part of the FoV, corresponding to directions for which the whole source flux recorded by the detector has been modulated by the mask pattern, is called Fully Coded FoV (FCFoV).  The Partially Coded FoV corresponds instead to directions for which only  part of the detected source flux has passed through the  mask pattern.  The sensitivity is nearly uniform in the FCFoV, while it gradually decreases in the PCFoV, as a result of the detector area used to record the source counts shrinking, as illustrated in the left panel of Fig.~\ref{fig:fov}. 

\begin{figure} [ht]
   \begin{center}
         \includegraphics[angle=90,height=6cm]{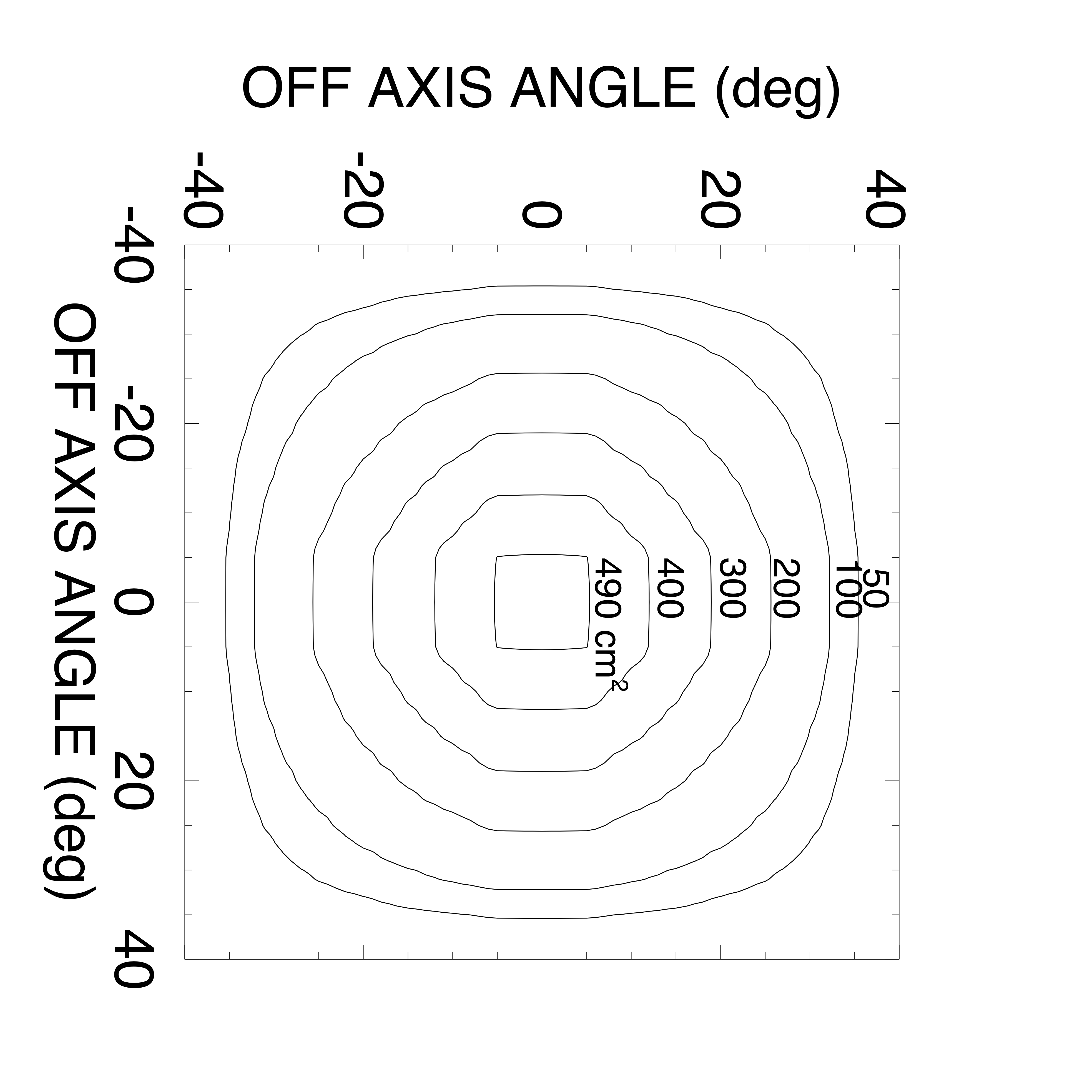}    
      \includegraphics[angle=90,height=5.4cm]{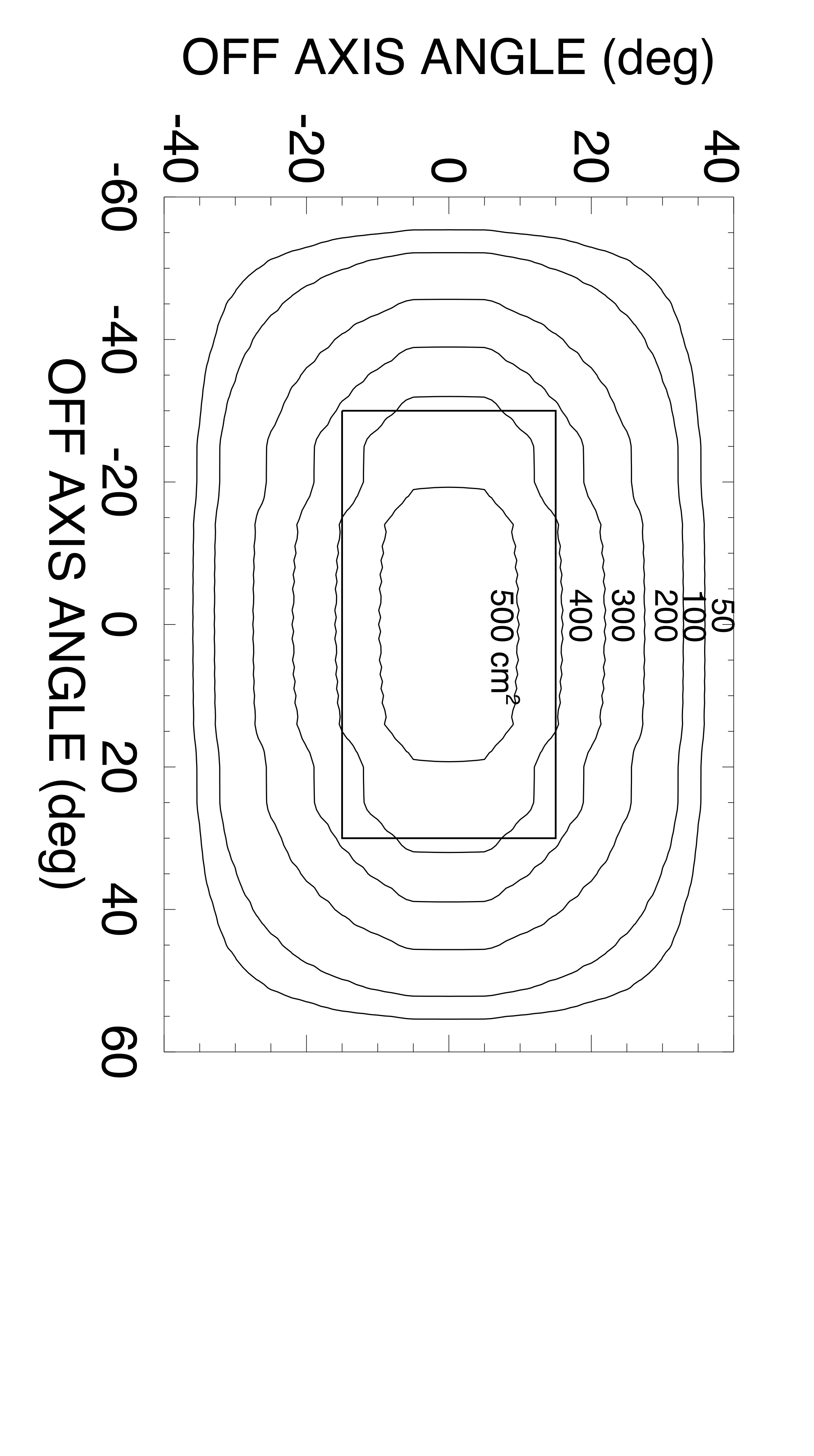}    
   \end{center}
   \caption[blabla] 
   { \label{fig:fov2} 
Effective area (at 10 keV) as a function of off-axis angle in a single XGIS camera (left panel) and in the sum of the two cameras (right panel). The effect of a mask with open fraction of 50\% is included.
}
   \end{figure}

The FCFoV and PCFoV of a single XGIS camera are concentric  squares with sides of  $2\arctan \displaystyle \frac{(D_M-D_D)/2}{H}   \sim  11^{\circ}$,  and  $2 \arctan \displaystyle \frac{(D_M+D_D)/2}{H}       \sim  77^{\circ}$, respectively.  The PCFoV corresponds to a solid angle of 1.6  sr. The variation of effective area as a function of the off-axis angle is shown in the left panel of Fig.~\ref{fig:fov2}.
Given that the two XGIS cameras   point at different directions, offset by  $\pm20^{\circ}$, their PCFoVs partially overlap and provide an overall rectangular field of view of  $\sim  77^{\circ}\times117^{\circ}$ = 2.24 sr.   Fig.~\ref{fig:fov2}  (right panel) shows the angular dependence of the effective area obtained by the combination of the  two cameras. The rectangle of $31^{\circ}\times61^{\circ}$ indicates the field of view of the SXI instrument. 
It can be noted that, by combining the two XGIS cameras, a relatively uniform effective area, larger than $\sim80\%$ of its maximum value, is obtained in the whole sky region that is covered in the soft energy range by the SXI   ($\sim$0.5 sr).

\subsection{Angular resolution and source location accuracy}
\label{sec:sla}  

For a given    spatial resolution of the detector (in the XGIS case this corresponds to the pitch $P=5$~mm),  the angular resolution ($\Theta$)   of a coded mask imaging system depends on  the mask distance and on the size of the mask elements ($M$). For on-axis sources it can be approximated by  $ \displaystyle \Theta \sim ((M/H)^2 + (P/H)^2)^{1/2} $.  The   value of $M$ has not been decided yet, since it will be optimized with a careful trade-off during the phase-B study, when further design elements will be defined. However, it can be anticipated that the ratio $M/P$ will likely be in the range 1.5--2.5,  leading to     $\Theta\sim 0.8-1.2^{\circ}$.

The source location accuracy  ($SLA$) is generally much better than the angular resolution, because it depends also on the statistical significance ($SNR$) of the  source detection. In particular, the location accuracy is given approximately by
$ \displaystyle  SLA = \lambda \Theta$ / $SNR$.     
The factor $\lambda$, of the order of a few, depends on how the $SLA$ and source significance are defined, as well as on the imaging properties of the mask, that in general are not those of an ideal system. 
In the following, we consider for the $SLA$  the 90\% containment radius, $R_{90}$  (i.e. the angular distance from the derived coordinates that has a 90\% probability of containing the true position of the source) and we  measure the source significance by  the ratio between the number of source counts and the square root of the number of  background counts,  $SNR= S / \sqrt{B}$.   
By performing Monte Carlo simulations with different mask patterns and adopting a representative value $M$=9 mm, we obtained a preliminary  estimate of  $\lambda\sim$2.5.

Note that all the definitions and quantities in this subsection refer to a single XGIS camera, unless specified differently. 
For the sources detected by both cameras it will be possible to improve the localization accuracy by combining the two data sets.

   \begin{figure} [ht]
   \begin{center}
      \includegraphics[angle=90,height=6cm]{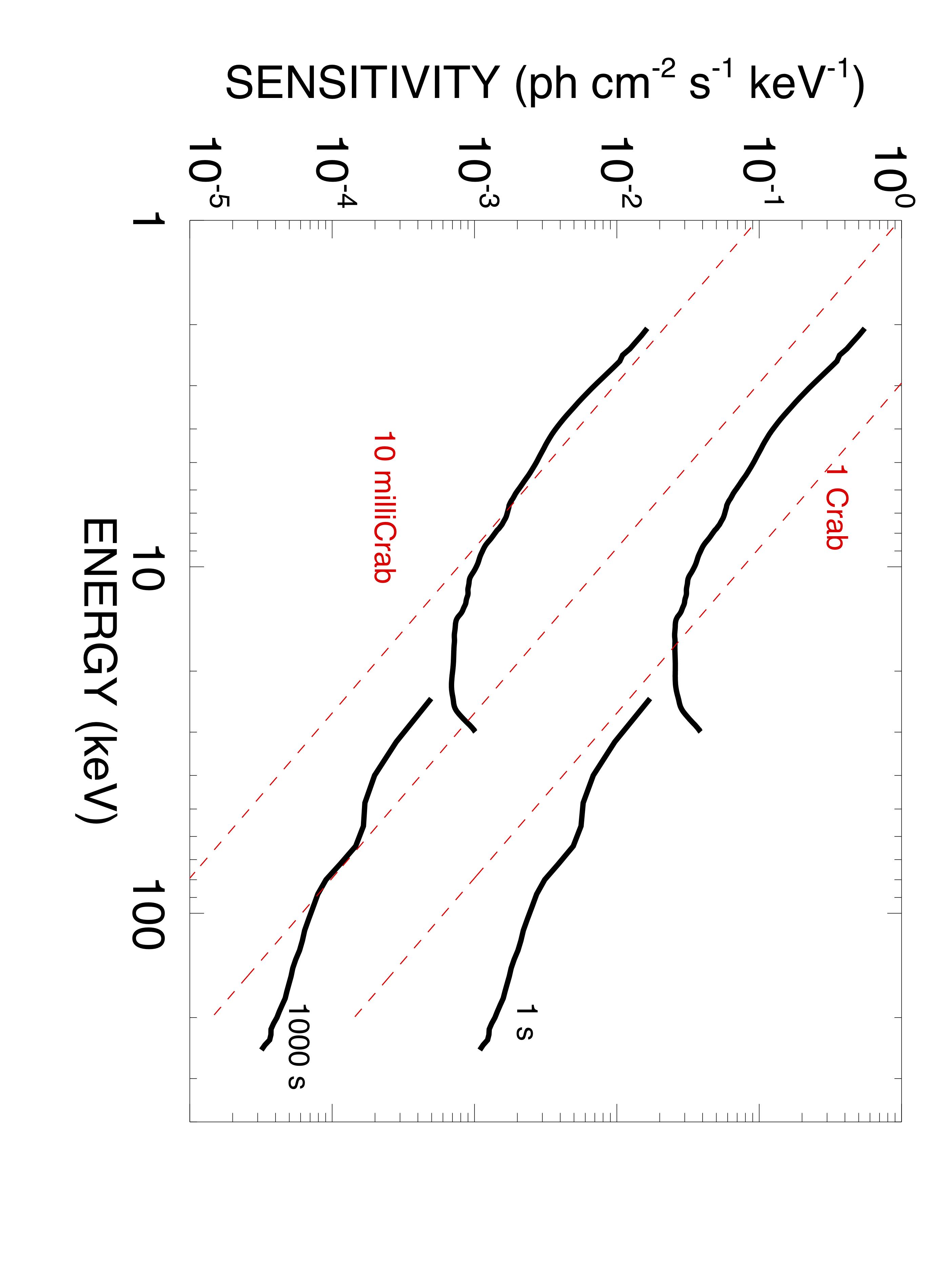}
      \includegraphics[angle=0,height=6cm]{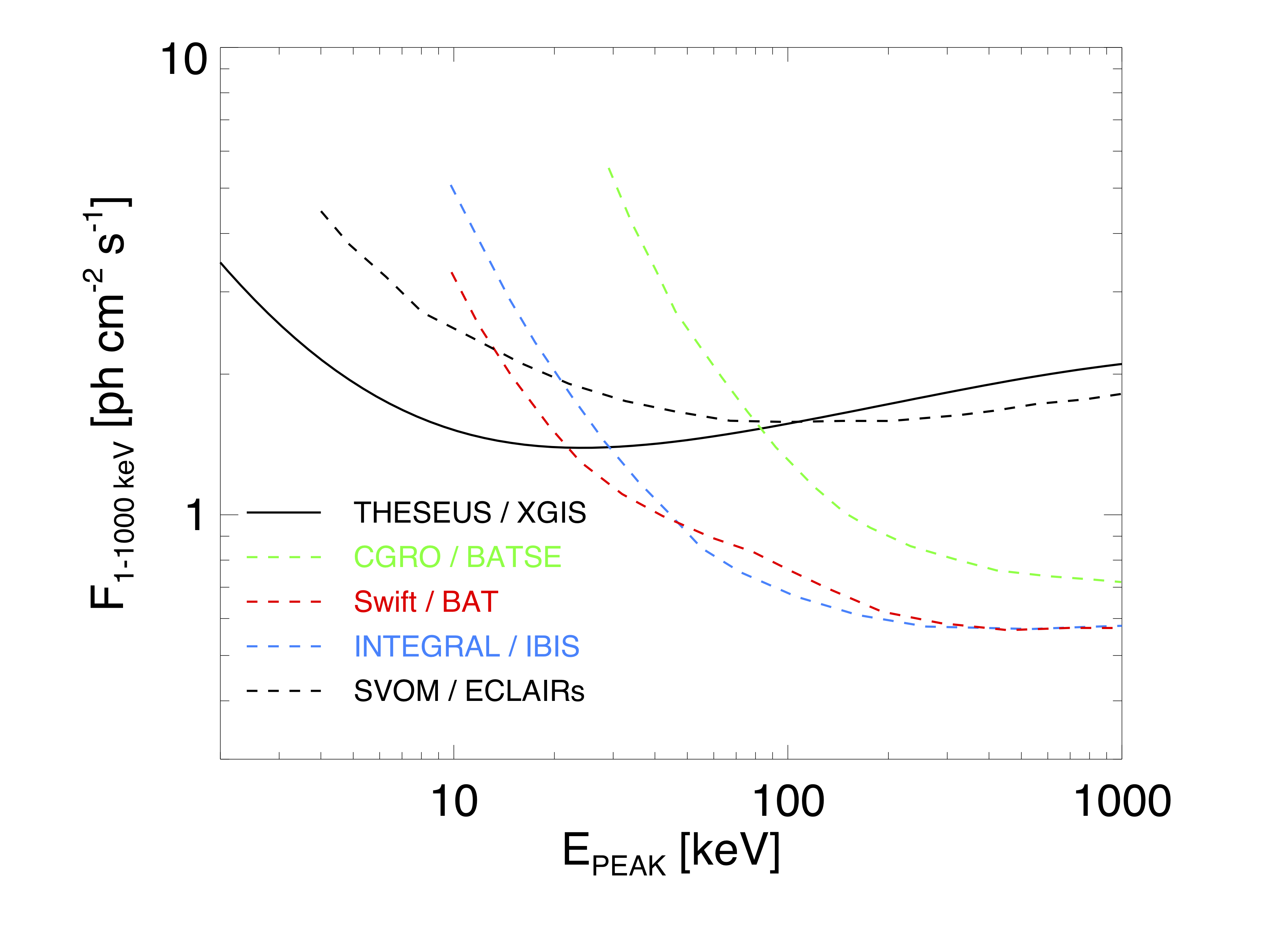}    
   \end{center}
   \caption[blabla2] 
   { \label{fig:sens} 
  Left: Differential flux sensitivity in the Fully Coded Field of view of one XGIS camera  (5$\sigma$, $\Delta E / E=1$).
  Right: Integral (1-1000 keV)  flux  sensitivity of the XGIS  as a function of the GRB peak energy. The XGIS curve refers to   the SDD sensitivity in the 2-30 keV range and is compared to that of  instruments on other satellites (adapted from \cite{schanne09}).  A Band spectrum with $\alpha=-1$ and $\beta=-3$, and an integration time of 1 s have been assumed.}
   \end{figure}

   \begin{figure} [ht]
   \begin{center}
      \includegraphics[angle=0,height=6cm]{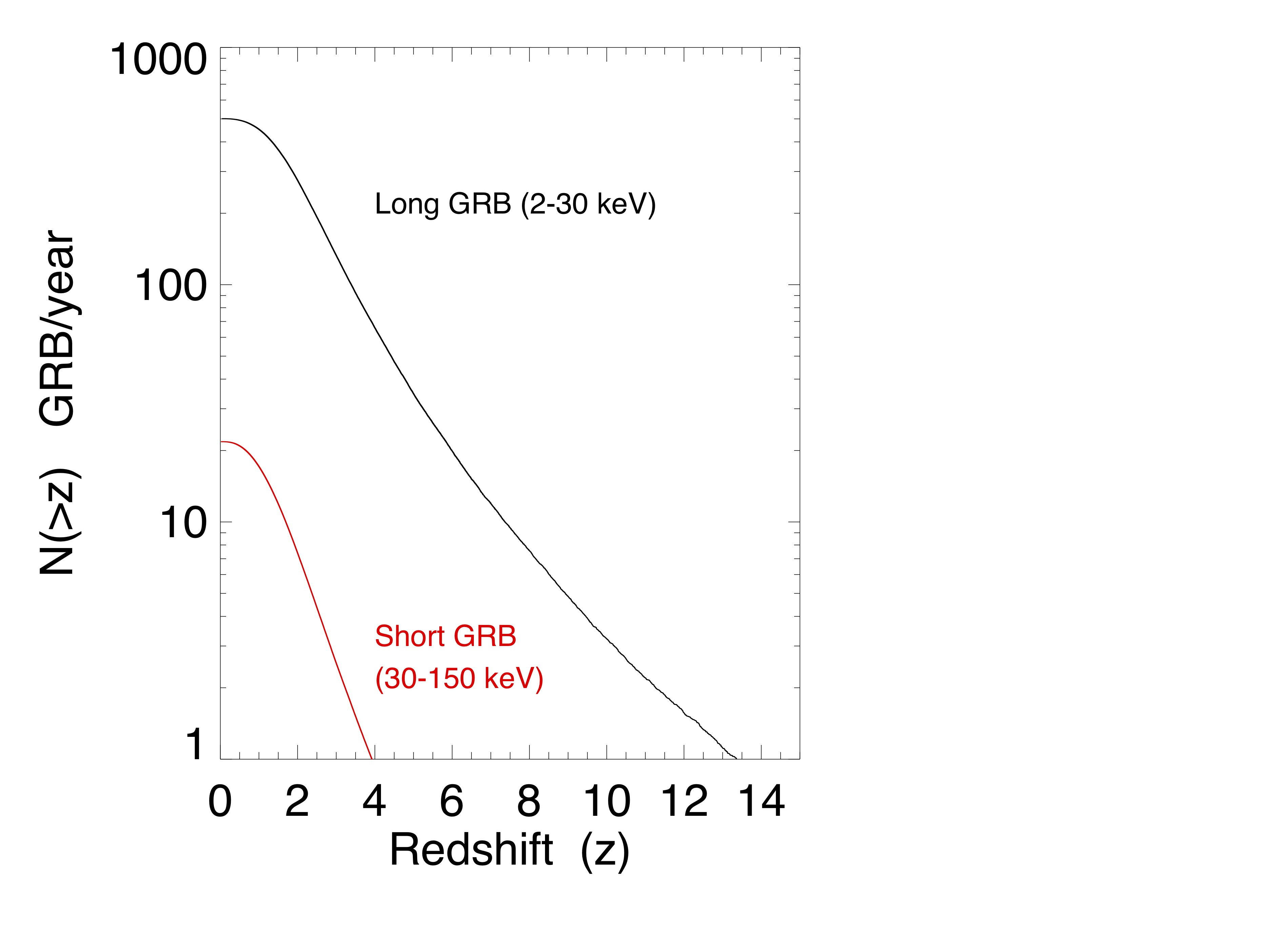}    
   \end{center}
   \caption[bla] 
   { \label{fig:rate} 
 Expected rates of GRB detected by the XGIS above a   threshold $SNR$=7 (the rates are not corrected for the observing efficiency, assumed here of 100\%. }
   \end{figure} 
   
\subsection{Effective area and imaging sensitivity}
\label{sec:nifov}  

The detection plane of one XGIS unit has an active geometric area of 6400$\times0.45^2= 1296$ cm$^2$, for what concerns the CsI(Tl) scintillators.   The individual SDD pixels have an active cross section smaller than that of the underlying scintillators, due to partial covering by the PCB 
 (3.5$\times$5.0 mm$^2$ for 4800 pixels and 3.5$\times$3.5 mm$^2$ for the remaining ones). Therefore, the active geometric area of the SDD is of 1036 cm$^2$.

The effective area (one unit) useful for imaging is shown in the right panel of Fig.~\ref{fig:fov} as a function of energy and for different off-axis angles. A  mask with open fraction of 50\% has been assumed.  Using the expected XGIS  background\cite{campana20}, 
we derive the  sensitivity   as a function of energy  shown in the left panel of Fig.~\ref{fig:sens} (one unit).  This sensitivity has been computed assuming a perfect mask (i.e. with  transmission $\eta_O$=1 and $\eta_C$=0 for the open and closed elements, respectively), which is a good approximation in the imaging energy ranges used below (2-30 keV for the SDD detectors, 30-150 keV for the scintillators). For each unit, the sensitivity is nearly uniform in the FCFoV and then it gradually decreases at larger off-axis angles.  

Fig.~\ref{fig:sens} (right panel) shows the XGIS sensitivity  as a function of the GRB spectral hardness. This is parametrized by the peak energy $E_p$ (see below).  It can be seen that, thanks to the broad energy range extending down to 2 keV, the XGIS instrument has a better sensitivity for the softer GRBs, compared to other instruments.

   \begin{figure} [ht]
      \includegraphics[angle=90,height=6.5cm]{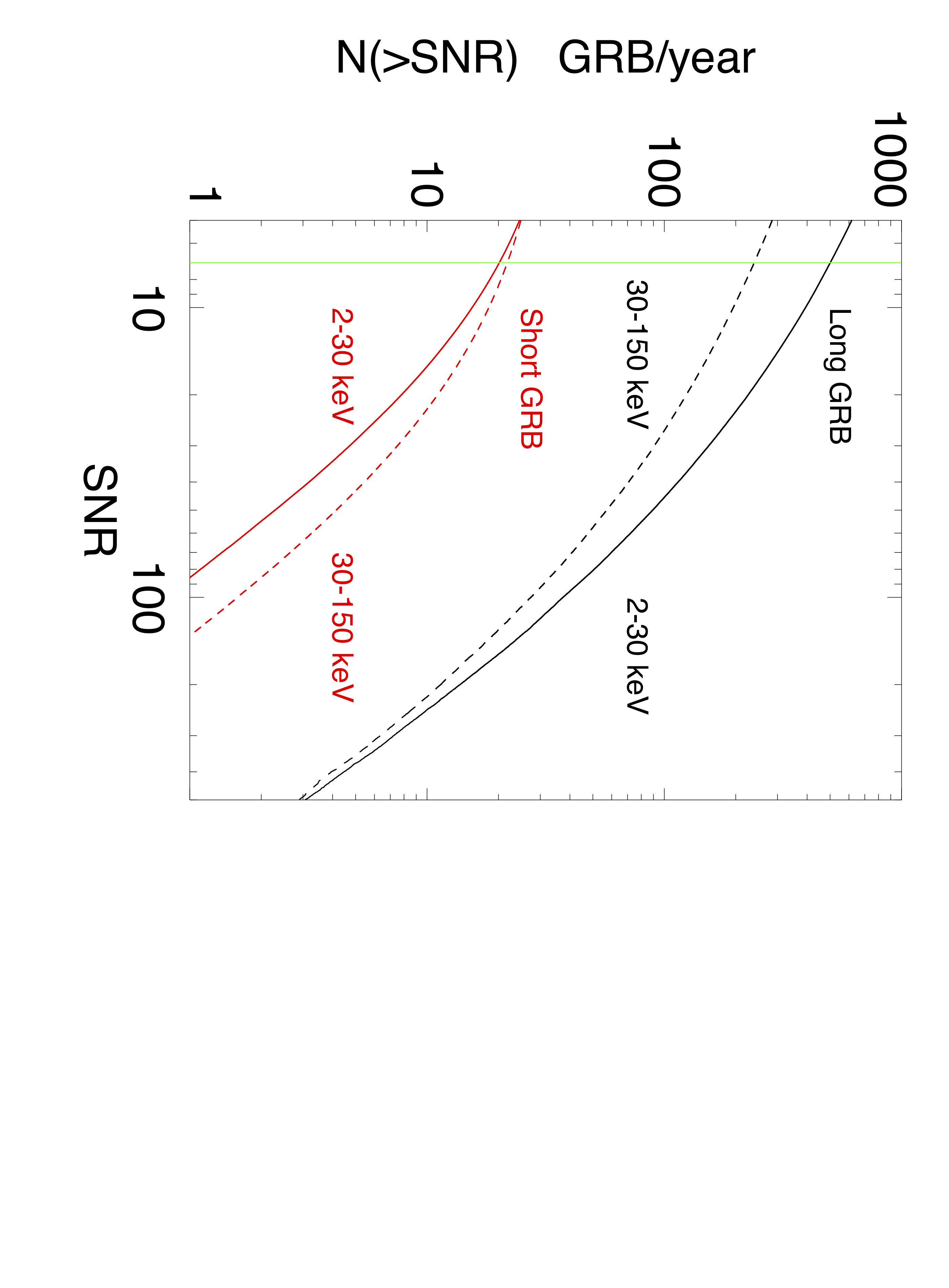}    
        \includegraphics[angle=90,height=6.5cm]{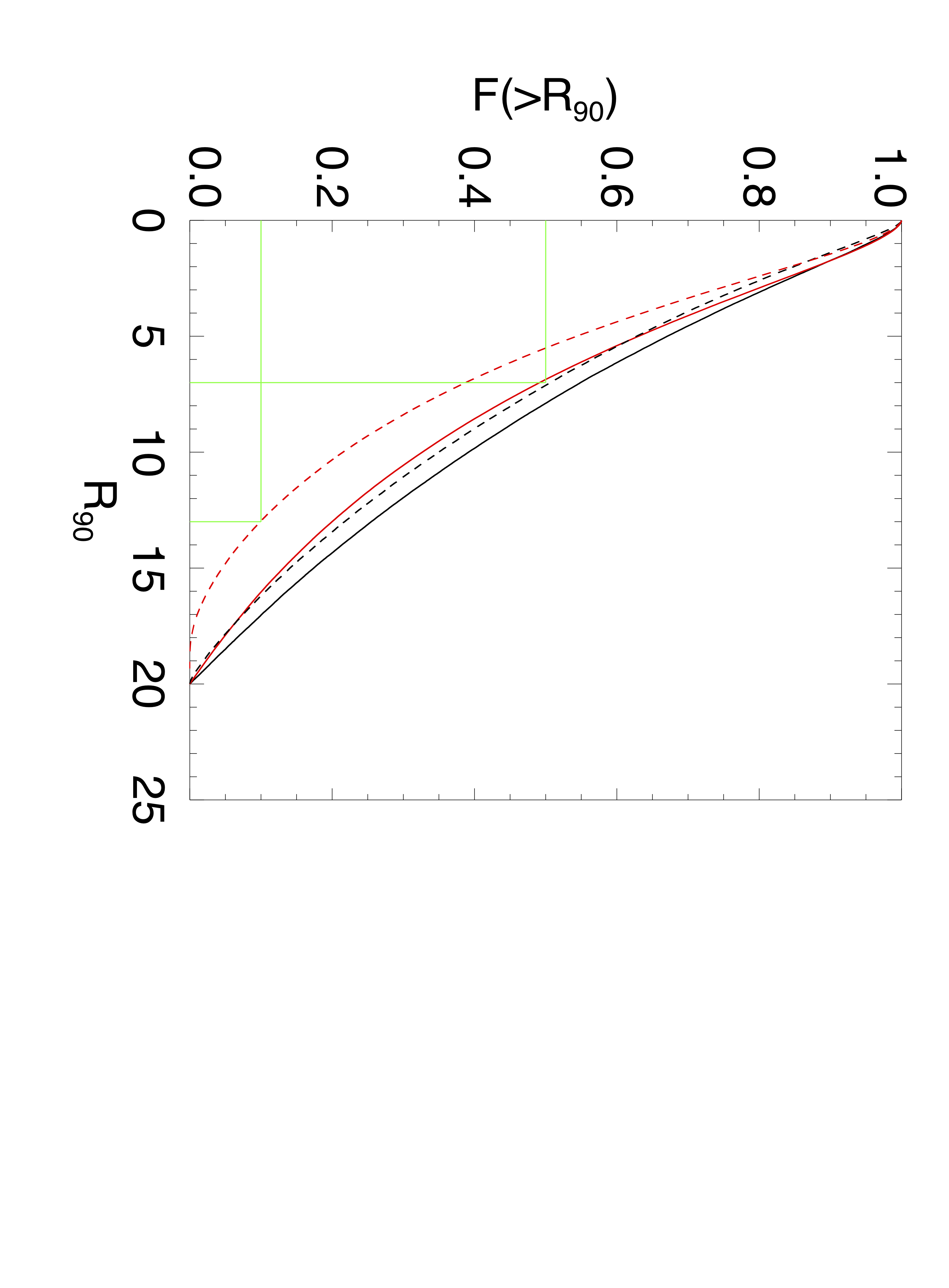} 
   \caption[bla] 
   { \label{fig:snr} 
 Left: Integral distribution of signal to noise ratio of the short and long GRBs. The green vertical line indicates the adopted threshold $SNR$=7.
 Right: Normalized integral distribution of the 90\% c.l. localization uncertainty for long (black) and short (red) GRBs detected with $SNR>7$. The solid lines refer to the localizations obtained using only the SDD  (2-30 keV) or the CsI (30-150 keV) detectors for the long or short GRBs, respectively. The dashed lines refer to the improved localizations that can be obtained when the  data of both detectors are combined.  The green lines show that 50\% of the detected long GRBs have R$_{90}<7'$ and   90\% of the short GRBs have R$_{90}<13'$. }
   \end{figure}

\section{Simulations with GRB populations}
\label{sec:pop}  

In order to evaluate the  XGIS performances  for the study of GRBs, we used two   populations of simulated sources:  one for the GRBs of the ``long'' class,  resulting from core collapse supernovae and one for  those of the  ``short''  class, originating in the coalescence of  binaries composed of neutron stars or black holes.
These   state-of-the-art GRB populations have been computed\cite{ghi15,ghi16} with a set of assumptions (luminosity function and redshift distribution, spectral parameter distributions for a typical Band spectrum\cite{band93}, correlations between the prompt emission spectral peak energy and total energy/luminosity) which describe their intrinsic properties. The free parameters of these distributions have been optimized by reproducing the observed properties of the populations of GRBs detected by Fermi and Swift. 
Both synthetic populations contain the  spectral parameters (peak energy $E_p$, photon index at low energy, $\alpha$, and at high energy, $\beta$), luminosity, duration and redshift of  two millions  GRBs. Their positions in the XGIS field of view are assigned randomly in each simulation with an isotropic   distribution.  To derive annual detection rates in the XGIS, the populations have been normalized to the values observed by Swift/BAT for the long GRBs and by Fermi/GBM for the short GRBs.

To compute the expected rate of GRBs detected by the XGIS we perform the following steps. We first assign random sky positions to all the GRBs and convert them to instrumental coordinates in the two XGIS units. Then, for each GRB, we compute the number of counts, $S_i$ detected in each unit ($i=1,2$) during an integration time T$_{EFF} = (1+z) E_{ISO} / L_{ISO}$,  where $z$, $E_{ISO}$ and  $L_{ISO}$ are the redshift, isotropic energy and isotropic luminosity of the GRB.  This is done by an energy integration of the burst spectrum multiplied by the effective area corresponding to the appropriate instrumental coordinates.
Similarly we compute the   background counts, $B_i$, in the same time and energy intervals,   and   $SNR_i = S_i / \sqrt{B_i}$.  Finally, we derive the total significance of the detection as  $SNR = (SNR_1^2+SNR_2^2)^{1/2}$. 
The sky region imaged by both cameras corresponds to $30\%$ of the total XGIS FoV, but due to the off-axis dependence of the sensitivity, about 40\% of the   GRBs detected by the XGIS are in this region.

In Fig.~\ref{fig:rate} we show the redshift distribution of the long and short GRBs detected above a threshold $SNR$=7 in two representative energy ranges.  These rates represent lower limits because only two fixed energy ranges have been used. In practice, the on board GRB trigger algorithms will operate simultaneously in several different energy ranges, thus  providing a better sensitivity for different spectral shapes.  
Furthermore, these rates consider the population of GRBs whose jet is aligned with the line of sight. In the local Universe this rate could be increased by the detection of slightly off axis events, particularly relevant as possible electromagnetic counterparts of gravitational waves\cite{str18}.
 Short GRBs have on average harder spectra than the long ones\cite{ghi04} and are   best detected in the 30-150 keV range by the CsI scintillators, while the opposite is true for the long GRBs. This can be seen in the left panel of Fig.~\ref{fig:snr} that shows the distributions of $SNR$ derived in the two detectors.  Note that all the rates plotted in  Figs.~\ref{fig:rate} and \ref{fig:snr} refer to an observing efficiency of 100\%. Therefore, they must be corrected for various effects, depending on the instrument and satellite operations,  in order to estimate the actual number of   GRBs that will be detected by the XGIS.

From the $SNR$ values shown in Fig.~\ref{fig:snr} it is possible to estimate the distribution of source location accuracy of the detected GRBs. For a mask with $M$=9 mm  ($\Theta$=56$'$) and $\lambda$=2.5, as derived from imaging simulations, we obtain the distributions of the 90\% error radius $R_{90}$ shown in the right panel of  Fig.~\ref{fig:snr} for the long (black line) and short (red lines) GRBs. As an example, the green lines show that 50\% of the detected long GRBs have R$_{90}<7'$ and   90\% of the short GRBs have R$_{90}<13'$.

\acknowledgments 
 We acknowledge the financial support of the Italian Space Agency and of the Italian National Institute of Astrophysics through the ASI-INAF Agreement n. 2018-29-HH.0.

\bibliographystyle{spiebib} 

\end{document}